\documentclass[preprint,showpacs,preprintnumbers,amsmath,amssymb]{revtex4}

\usepackage{graphicx}
\usepackage{dcolumn}
\usepackage{bm}

\begin{document}

\title{\large Critical current measurements in superconductor - ferromagnet - superconductor junctions of
$YBa_2Cu_3O_y$-$SrRuO_3$-$YBa_2Cu_3O_y$:\\
No evidence for a dominant proximity induced triplet
superconductivity in the ferromagnetic barrier}

\author{G. Koren}
\email{gkoren@physics.technion.ac.il} \affiliation{Physics
Department, Technion - Israel Institute of Technology Haifa,
32000, ISRAEL} \homepage{http://physics.technion.ac.il/~gkoren}

\author{T. Kirzhner}
\affiliation{Physics Department, Technion - Israel Institute of
Technology Haifa, 32000, ISRAEL}

\author{P. Aronov}
\affiliation{Physics Department, Technion - Israel Institute of
Technology Haifa, 32000, ISRAEL}

\date{\today}
\def\bfig {\begin{figure}[tbhp] \centering}
\def\efig {\end{figure}}

\normalsize \baselineskip=8mm  \vspace{15mm}


\begin{abstract}

Transport measurements in ramp-type junctions of
$YBa_2Cu_3O_y-SrRuO_3-YBa_2Cu_3O_y$ with $T_c$ values of either 80-90 K or 60-70 K are
reported. In both type of junctions but without a barrier ("shorts"), the supercurrent densities at 4.2 K reached  7.5 and 3.5 MA/cm$^2$, respectively, indicating the high quality of the fabrication process. Plots of the critical current versus thickness of the ferromagnetic barrier at 4.2 K show exponential decays with decay lengths of 1.1 nm for the 90 K phase and 1.4 nm for the 60 K phase, which are much shorter than the relevant coherence lengths $\xi_F\sim 5-6$ nm or $\xi_N\sim$16 nm of $SrRuO_3$. We thus conclude that there is no dominant proximity induced triplet superconductivity in the ferromagnet in our junctions.

\end{abstract}

\maketitle

In a recent feature article in the January 2011 issue of Physics Today, Matthias Eschrig reviews the rapidly growing field of proximity induced triplet superconductivity (PITS) in ferromagnets in contact with a superconductor which attracted much attention in the past few years \cite{PT}. In the present study, we report on a null result of this effect in superconductor - ferromagnet - superconductor (SFS) junctions of the high temperature superconductor $YBa_2Cu_3O_y$ (YBCO) and the itinerant ferromagnet $SrRuO_3$ (SRO), and believe that this new observation should help refine future PITS theories. Basically, standard singlet superconductivity and strong ferromagnetism are two antagonistic phenomena due to their different spin ordering configurations. It should therefore be hard to obtain supercurrents in SFS junctions when the barrier thickness $d_F$ is much larger than the short coherence length of the ferromagnet, either $\xi_F=\hbar v_F/2E_{ex}$ in the clean limit or  $\xi_F=\sqrt{\hbar D/2E_{ex}}$ in the dirty limit, which are affected mostly by the exchange energy $E_{ex}\sim k_B T_{Curie}$ \cite{RMPBergeret,RMPBuzdin}. This however is not the case if the singlet pairs in S, in the vicinity of the SF interface, would induce equal-spin triplet pairs in the ferromagnet via the proximity effect. Then, due to the compatibility of the triplet and ferromagnetic orders, a supercurrent could be maintained at low temperature over a long range of $d_F\sim 2\xi_N$ with a coherence length $\xi_N=\sqrt{\hbar D/2\pi k_B T}$ rather than the shorter $\xi_F$ ones.\\

A number of theoretical studies had predicted the PITS effect
which can originate in interface inhomogeneities
such as domain walls or spin mixing and spin flip scattering
\cite{Bergeret,Fominov,Eschrig,KB,VE}. Supercurrents were also
observed experimentally in SFS junctions with remarkably long
half-metal ferromagnetic $CrO_2$ barriers ($d_F=300-700\,nm$) and low $T_c$ s-wave
superconductors \cite{Gupta,Anwar}, and in highly polarized
$La_{2/3}Sr_{1/3}MnO_3$ (LSMO) barrier ($d_F=20\,nm$) and a high $T_c$ d-wave
superconductor \cite{Ivanov}. In the low $T_c$ case, the
supercurrent density ranged between 1-100 $kA/cm^2$ at
2 K, while in the high $T_c$ case, it was 5 $kA/cm^2$
at 4 K. In both cases however, no systematic measurements of the supercurrent were done versus the ferromagnetic barrier thickness. In the present study we do report on such measurements in SFS ramp type junctions of YBCO and SRO. At 4.2 K, the critical current plots versus the barrier thickness show  decay lengths $\xi$ which are much shorter than $\xi_F$ thus excluding the possibility of a dominant PITS component in the ferromagnet in our junctions.\\

\begin{figure} \hspace{-20mm}
\includegraphics[height=9cm,width=13cm]{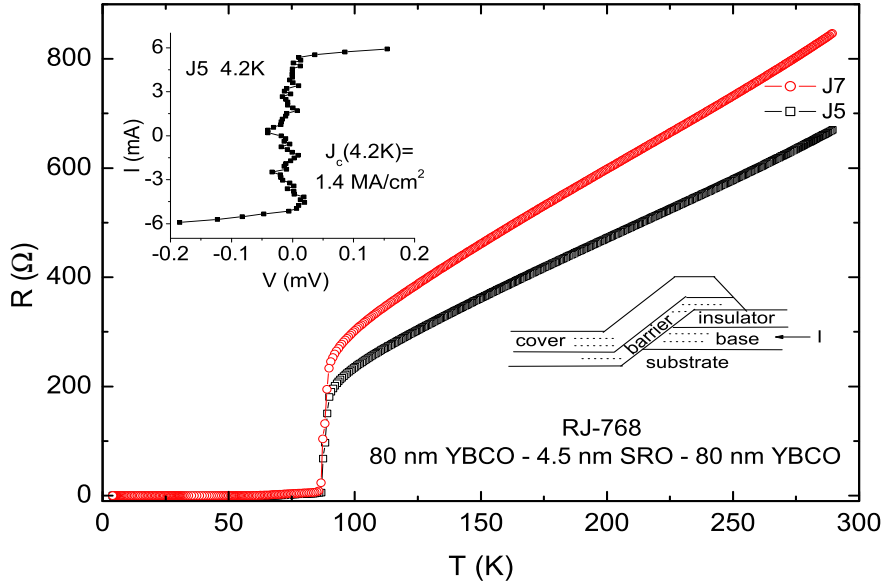}
\vspace{-0mm} \caption{\label{fig:epsart}Resistance versus
temperature of two YBCO/4.5nm SRO/YBCO junctions with the I-V curve of J5 at 4.2 K in the top inset. The
bottom inset shows a schematic drawing of a ramp-type junction,
where the 80 nm thick base and cover electrodes are made of YBCO
and the barrier is made of SRO.}
\end{figure}

About two hundred junctions on twenty wafers of (100) $SrTiO_3$ (STO) were prepared and characterized in the present study. For this, we fabricated ten ramp-type junctions on each wafer in the geometry shown in the inset to Fig. 1 by a multi step process \cite{Nesher123}. The junctions generally had different SRO barrier thickness ($d$
of 0, 4.5, 9, 13, 18, 22.5 and 45 nm) but the same YBCO electrodes' thickness (80 nm). They were fully epitaxial with the c-axis normal to the wafer, coupled in the a-b planes between the base and cover electrodes, oriented along the (100) STO direction,  had 5 $\mu m$ width and their cross section area was 0.4 $\mu m^2$. The resulting junctions were generally annealed under 50 Torr of oxygen pressure to produce optimally doped electrodes of the 90 K YBCO phase. Some junctions however, were re-annealed under oxygen flow of 0.1 Torr which yielded the 60 K phase of YBCO. The SRO barrier has remained unchanged under these annealing conditions. This allowed us to test if our critical current results are sensitive to the doping of YBCO and in what way.\\

Typical 4-probe results of the resistance versus temperature of two junctions with $d=4.5$ nm are shown in Fig. 1. In addition to the superconducting transition temperatures of the YBCO electrodes at 87-90 K, there are two weakly resistive tails of a few $\Omega$ down to 70 and 60 K where the junctions reach zero resistance. The top inset of this figure shows a current versus voltage  (I-V) curve at 4.2 K of the junction with the lower normal resistance and $T_c(R=0)=70$ K (J5). One can see that the critical current ($I_c$) is of about 5.5 mA while the critical current density ($J_c$) is 1.4 MA/cm$^2$. The noise near zero bias is due to bad contacts in this case, though generally gold coated contacts were used which had much lower noise. The second junction (J7) with the higher normal resistance and lower $T_c(R=0)=60$ K, had about half of the supercurrent of the first junction. This strong variation in the supercurrents is typical of our junctions with the ferromagnetic SRO barrier, and this effect becomes even more pronounced with increasing barrier thickness. With the present small barrier thickness of 4.5 nm however, we can not rule out the existence of microshorts due to the nanometer roughness of the two SF interfaces of the junctions \cite{Koren}, and the possibly incomplete coverage of the base electrode with the thin SRO layer. For comparison we fabricated and tested "shorts" which are ramp type junctions prepared by exactly the same process but without the barrier. These had at 4.2 K maximal $J_c$ values of 7.5 and 3.5 MA/cm$^2$ for the junctions with $T_c\sim$80-90 and $T_c\sim$60-70 K, respectively. Thus, the effect of possible microshorts in the junctions of Fig. 1 is not dominant, as their maximal supercurrent density is still a factor of about five (7.5/1.4) lower than that of the corresponding "short".\\

\begin{figure} \hspace{-20mm}
\includegraphics[height=9cm,width=13cm]{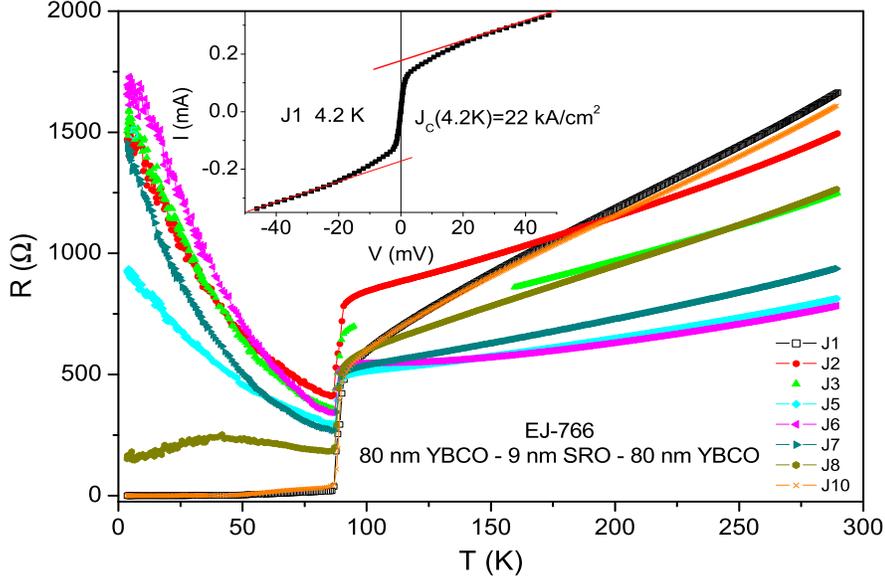}
\vspace{-0mm} \caption{\label{fig:epsart}Resistance versus
temperature of eight YBCO/9nm SRO/YBCO junctions with the I-V curve of one of the junctions with zero resistance at 4.2 K (J1) shown in the inset. }
\end{figure}

\begin{figure} \hspace{-20mm}
\includegraphics[height=9cm,width=13cm]{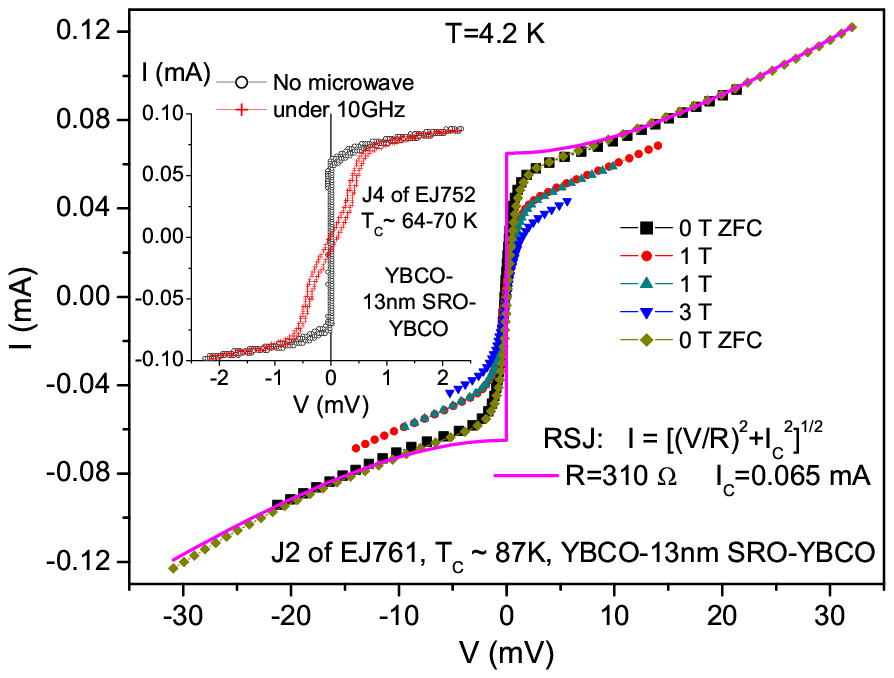}
\vspace{-0mm} \caption{\label{fig:epsart}Current versus voltage curves at 4.2 K of an
YBCO/13nm SRO/YBCO junction with RSJ behavior, under zero field cooling (ZFC) and under 1 and 3 T magnetic fields. The solid curve is an RSJ fit with the formula and parameters given in the figure. The inset shows two oscilloscope traces of I-V curves at 4.2 K near zero bias with and without 10 GHz microwave radiation, on a similar junction but on a different wafer with $T_c\sim$67 K.   }
\end{figure}

Fig. 2 shows the resistance versus temperature results of eight junctions with SRO barrier thickness of $d$=9 nm and $T_c\sim$87-90 K. The different normal state resistances are due to different lengths of the leads to the junctions. Actually, J5 and J6 have similar lead's lengths and therefore their normal resistances are quite similar. The same is true for J1 and J10, and also for J3 and J8. Fig. 2 clearly shows the wide spread of the junctions' resistance below $T_c$ of the electrodes. This effect has been observed before \cite{Aronov} and was attributed to the nonuniform interface resistance whose origin is still unclear. In the present study however, we shall not focus on the highly resistive junctions, but on those with lowest resistance which generally have the highest supercurrents. A typical I-V curve at 4.2 K of one of the junctions with zero resistance (J1) is shown in the inset to Fig. 2. One can see that the junction becomes slightly resistive with a resistance of a few $\Omega$ at a relatively low bias. This
generally depends on earlier magnetic field exposure or history of the junction (trapped flux), or on the intrinsic magnetic field emanating from the ferromagnetic barrier. Both of these effects lead to flux creep resistance with increasing bias. At higher bias, the critical current is reached and a change to the normal state is observed where the rounding is now due to flux flow and also to thermal noise. The high bias slope of the I-V curve yields a normal resistance $R_N$  of $\sim 300\,\Omega$ which is much higher than that calculated from the SRO resistivity and junction geometry (about 10 $m\Omega$). This result therefore, originates in the the two interfaces of the junctions as has already been observed before \cite{Aronov}. We generally determine the $I_c$ values of the junctions by extrapolating the high bias data to zero bias as shown in the inset to Fig. 2. At 4.2 K this yields  a $J_c$ value of 22 kA/cm$^2$ for junction J1 of Fig. 2.  Junction J10 had comparable supercurrent density, while the other junctions had much smaller critical currents or not at all. We therefore decided that for comparison between junctions with different barrier thickness we shall always take the maximal $I_c$  values of one or two junctions on each wafer.\\

The main panel of Fig. 3 presents I-V curves with a resistively shunted junction (RSJ) behavior at 4.2 K of a junction with an SRO barrier thickness of 13 nm and $T_c$ in the range of 85-89 K. The critical current can be determined by the use of the RSJ formula given in this figure, or by the extrapolation procedure as shown before in the inset to Fig. 2.  The extrapolation procedure however, underestimates the supercurrent in this case and we therefore chose to use the $I_c$ values derived from the RSJ formula.  Also shown in this figure are I-V curves under magnetic fields of 1 and 3 T, where the $I_c$ values are suppressed, the flux flow resistance increases, and the RSJ behavior is almost washed out.
The inset to Fig. 3 shows two oscilloscope traces of I-V curves with a zoom up on the low bias regime. These were measured on a junction with the same barrier thickness of 13 nm but on a different wafer that had $T_c$ values in the range of 64-70 K. Under microwave radiation, the junction became resistive at zero bias with a resistance of a few ohms. Unlike the dc measured results of the main panel which generally took 1-2 min. to record, the ac measured results in the inset were obtained with an averaging digital oscilloscope and took about 1 s. In the ac case, except for some hysteresis, no flux creep resistance could be observed up to the critical current at about 0.072 mA which is comparable to the result of the main panel. Since flux creep is more probable  in the dc measurements than in the ac ones due to the longer time available for de-pinning, this leads to the observed small flux creep resistance at low bias in the main panel which is absent in the inset. We note that the I-V curve without microwave radiation in the inset of Fig. 3 is similar to those obtained in Ref. \cite{Char} on similar junctions, although the normal resistance here is much larger ($\sim$100 $\Omega$).\\

 \begin{figure} \hspace{-20mm}
\includegraphics[height=9cm,width=13cm]{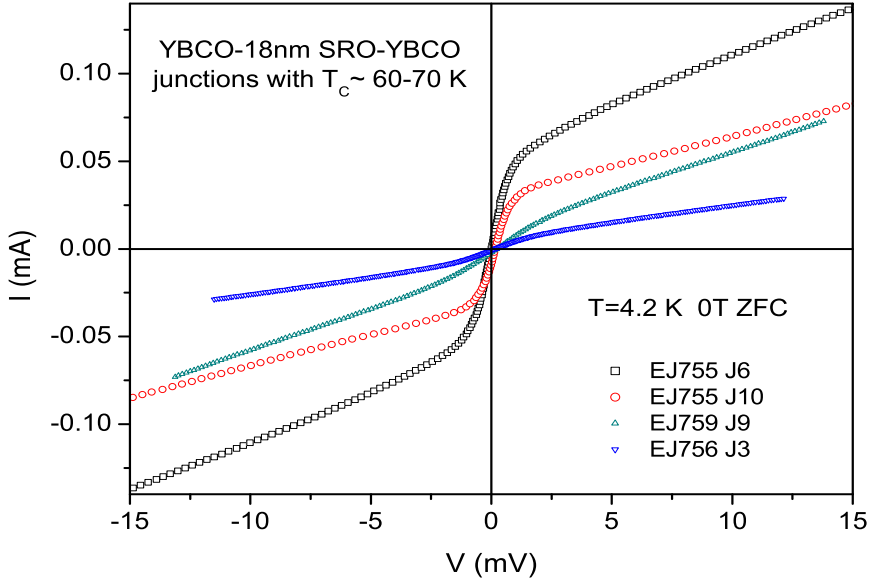}
\vspace{-0mm} \caption{\label{fig:epsart} Current versus voltage curves at 4.2 K of four
YBCO/18nm SRO/YBCO junctions with $T_c\sim$ 60-70 K and with the highest $I_c$ values on three wafers.}
\end{figure}

Fig. 4 shows I-V curves at 4.2 K of underdoped junctions with $T_c\sim$ 60-70 K and 18 nm SRO barrier thickness on three different wafers. The spread of the maximal $I_c$ values here is quite large and ranges between 7 and 60 $\mu A$. The low bias resistance due to flux flow is quite pronounced and it seems hard to distinguish between a critical current and a zero bias conductance peak (ZBCP) due to bound states. ZBCP were observed before in the same kind of SFS and SF junctions \cite{Aronov,Tal}. In SF junctions however, where no supercurrent exists the normalized ZBCP are small, typically up to 0.1. Since the normalized conductances $dI/dV$ at zero bias in Fig. 4 here range between 3 and 20, we conclude that the apparent ZBCP contribution to the critical current is negligible.\\

\begin{figure} \hspace{-20mm}
\includegraphics[height=9cm,width=13cm]{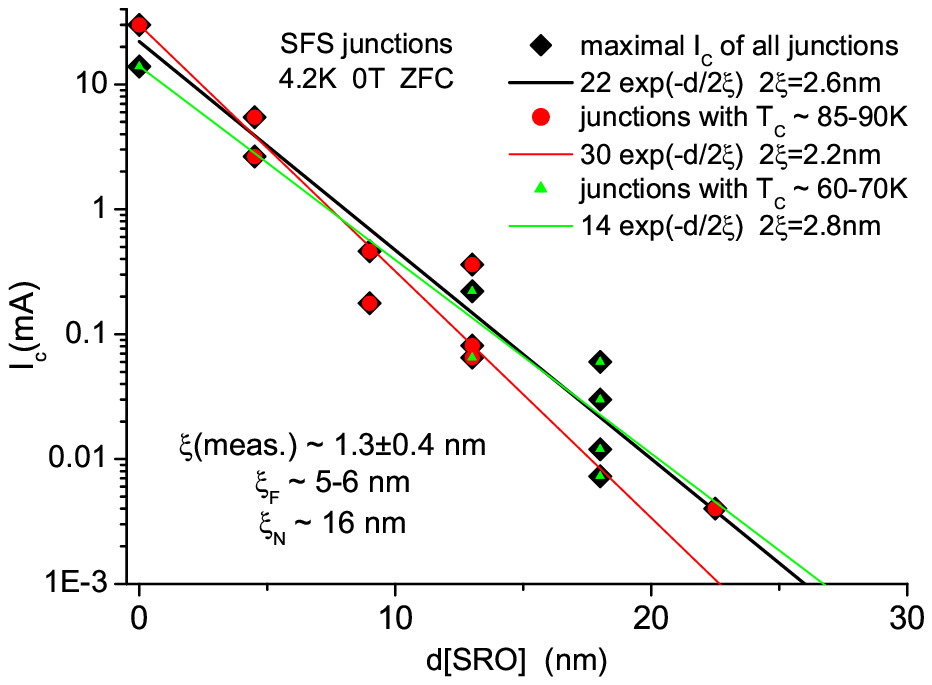}
\vspace{-0mm} \caption{\label{fig:epsart} Maximal critical current values at 4.2 K of one or two junctions on each wafer as a function of the SRO barrier thickness. Data is shown for all junctions (diamonds), the junctions with $T_c\sim80-90$ K (circles) and the junctions with $T_c\sim60-70$ K (triangles), with the corresponding exponential decay fits.  }
\end{figure}

Fig. 5 presents the main result of the present study. It shows all the maximal measured $I_c$ values of one or two junctions on each wafer as a function of the SRO barrier thickness. It also shows which data point belongs to the 80-90 K phase and which to the 60-70 K phase of YBCO. No critical current could be found in the junctions with the 45 nm thick barrier. The three exponential decay fits of the data correspond to all data points, and to the two different YBCO phases separately. We stress that due to the two SF interfaces in each SFS junction, the decay length is $2\xi$ rather than $\xi$. The immediate clear result is that all the three decay lengths obtained from the fits $\xi$=1.1, 1.3 and 1.4 nm are significantly shorter than $\xi_F$ of SRO which is either about 4.8 nm in the clean limit where $\xi_F=\hbar v_F/2E_{ex}$, or about 6.2 nm in the dirty limit where $\xi_F=\sqrt{\hbar D/2E_{ex}}$. These values were obtained using a Fermi velocity $v_F\sim 2\times 10^5$ ms$^{-1}$ and a mean free path at 4 K of $\ell\sim$ 14 nm \cite{Allen,Santi}, with the diffusion coefficient $D=v_F\ell/3$. For the exchange energy we used $E_{ex}\sim k_B T_{Curie}\sim 13$ meV, which is quite close to the $\sim$10 meV value  obtained from Faraday rotation measurements \cite{Dodge}. The latter though depends on a large subtraction of phonon contribution, so we used the former. Our measured $2\xi$ values of 2.2-2.8 nm can be qualitatively compared with those obtained in Ref. \cite{Char}, where mixed data of $J_c$ in junctions with SRO as well as $CaRuO_3$ barriers yields $2\xi$=6.2 nm. This larger value is apparently affected also by the low $J_c$ at $d_F$=0 ("short") in their study, which is smaller by a factor of $\sim$20 than in the present work. If a significant amount of equal-spin triplet pairs are induced in the ferromagnetic SRO barrier, we should have actually had to compare the measured $\xi$ values with $\xi_N$ of SRO provided a single magnetic domain is involved, since then no spin flip scattering would occur. In SRO films with normal c-axis orientation, the domain walls are in the (110) direction and their spacing is of about 1000 nm \cite{Marshall}. Thus in the present (100) oriented junctions with up to 45 nm thick barriers, transport occurs mostly via single domains with very little scattering at domain walls.  In this case, one obtains $\xi_N=\sqrt{\hbar D/2\pi k_B T}\sim$16 nm  in the dirty limit of SRO which is obviously  much larger than either of the $\xi_F$ values given above. We thus conclude that in the present SFS junctions, no significant PITS affects the measured critical currents.\\

Finally, we shall check the present result in the context of previous results on the YBCO-SRO system \cite{Asulin,Aronov,Tal}. The scanning tunneling spectroscopy results of SRO/(100)YBCO bilayers show a long range penetration of the superconducting order parameter via the SRO layer up to $d_F=26$ nm \cite{Asulin}, but only along lines that are correlated with the magnetic domain wall structure of the ferromagnet. In SFS and SF junctions, ZBCP were found whose zero field magnitude also correlated with the number of domain walls in the SRO barrier \cite{Aronov,Tal}. Both results were interpreted as due to non-local crossed Andreev reflection effect (CARE) near domain walls crossing of the interface, but could also be partially attributed to PITS. Due to the small fraction of the junction cross section area where these effects can occur ($\xi_S\times d_F\times N$ where $\xi_S\sim 2$ nm is the coherence length of YBCO, $d_F$ is the SRO layer thickness and $N$ the number of domain walls crossing of the interfaces in the junctions), their total contribution to the critical current is apparently small. Therefore, the present critical current results with the very short $\xi$ values can be explained as due to dominant local Andreev reflections that are present in the partially spin-polarized SRO barrier ($P\sim$ 50\% at 4.2K \cite{Beasley}). We are currently checking the critical currents of YBCO-$La_{2/3}Ca_{1/3}MnO_3$-YBCO junctions, where the polarization of the manganite is almost 100\% and the domain walls are much broader. We expect that in these junctions, a more dominant contribution to the critical current by PITS will be observed.\\

In conclusion, very short decay lengths of the critical current on the order of 1-2 nm were observed in the ferromagnetic barrier of YBCO-SRO-YBCO junctions, which are much shorter than the corresponding penetration lengths $\xi_F$ and $\xi_N$ of SRO. This result is attributed to the absence of a dominant proximity induced triplet superconductivity in the SRO layer in the present junctions. Future PITS theories will have to account for this observation.\\

{\em Acknowledgments:}  We thank Emil Polturak for useful discussions. This research was supported in part by the Israel Science Foundation, the joint German-Israeli DIP project and the Karl Stoll Chair in advanced materials at the Technion.\\

\bibliography{AndDepBib.bib}

\bibliography{apssamp}

\end{document}